\documentclass[12pt]{article}
\usepackage{euler}
\usepackage{amsmath}
\usepackage{amssymb}
\usepackage{amstext}
\usepackage{amscd}
\usepackage{amsfonts}
\DeclareMathAlphabet{\EuFrak}{U}{euf}{m}{n}
\DeclareMathAlphabet{\EuScript}{U}{eus}{m}{n}

\title{{\bf Ultradistributions and The Fractionary
Schr\"{o}dinger Equation}
\thanks{\it{This work was partially supported by Consejo
Nacional de Investigaciones Cient\'{\i}ficas ;
Argentina.}}}
\author{A. L. De Paoli and M. C. Rocca\\
Departamento de F\'{\i}sica, Fac. de Ciencias Exactas,\\
Universidad Nacional de La Plata.\\
C.C. 67 (1900) La Plata. Argentina.}

\date{December 15, 2009}

\begin{document}

\maketitle

\vspace{-5mm}

\begin{abstract}

In this work, we generalize the results of Naber about the
Fractionary Schr\"{o}dinger Equation with the use of the theory
of Tempered Ultradistributions. Several examples of the
use of this theory are given. In particular we evaluate the
Green's function for a free particle in the general case.


\end{abstract}

\newpage

\renewcommand{\theequation}{\arabic{section}.\arabic{equation}}

\section{Introduction}

The properties of ultradistributions (ref.\cite{tp1,tp2}) are well
adapted for their use in fractional calculus. In this respect we have
shown that it is possible (ref.\cite{tt2}) to define a general
fractional calculus with the use of them.

Ultradistributions  have the
advantage of being representable by means of analytic functions.
So that, in general, they are easier to work with them.

They have interesting properties.One of those properties
is that Schwartz tempered distributions are canonical and continuously
injected into tempered ultradistributions and as a consequence the Rigged
Hilbert Space with tempered distributions is canonical and continuously
included in the Rigged Hilbert Space with tempered ultradistributions.

Fractional calculus has found motivations in a growing area concerning 
general stochastic phenomena.These include the appearance of alternative
diffusion mechanisms other than Brownian, as well as classical and quantum 
mechanics formalisms including dissipative forces, and therefore allowing 
an extension of the quantization schemes for non-conservative systems \cite{tg1}.
In particular it is interesting to study the fractional Schr\"{o}dinger equation. 
Our aim is to extend a previous study \cite{tt1} about this equation.
Using an analytical definition of fractional derivative \cite{tt2} we show here that 
it is possible to obtain a general solution for the time fractional equation,
for any complex value of the derivative index. Furthermore the associated
Green functions can be evaluated in a straightforward way.

This paper is organized as follow:\\
In section 2 we define the fractional Schr\"{o}dinger equation for all $\nu$
complex with the use of the fractional derivative defined via the theory
of tempered ultradistributions. In section 3 we solve this equation 
for the free particle and give three examples:$\nu=1/2,\nu=1$ and $\nu=2$.
In section 4 we realize the treatment of the potential well and we
analyze the cases $\nu=1/2,\nu=1$ and $\nu=2$. In section 5 we study
the Green fractional functions for the free particle in three cases:
the retarded Green function, the advanced Green function and the
Wheeler-Green function. As an example we prove that for $\nu=1$
these functions coincide with the Green functions of usual
Quantum Mechanics.In section 6 we discuss the results obtained
in the previous sections. Finally we have included 
three appendixes: a first appendix on distributions of exponential type, 
a second appendix on tempered ultradistributions and a third appendix
on fractional calculus using ultradistributions.

\section{The Fractional Schr\"{o}dinger Equation}

\setcounter{equation}{0}

Our starting point in the study of the fractional Schr\"{o}dinger equation is 
the current known Schr\"{o}dinger equation:
\begin{equation}
\label{ep5.1}
i\hbar{\partial}_t\psi(t,x)=-\frac {{\hbar}^2} {2m} {\partial}_x^2
\psi(t,x)+ V(x)\psi(t,x)
\end{equation}
According to ref.\cite{tt1}, (\ref{ep5.1}) can be writen as:
\begin{equation}
\label{ep5.2}
i T_p{\partial}_t\psi(t,x)=-\frac {L_p^2M_p} {2m} {\partial}_x^2
\psi(t,x)+ \frac {V(x)} {E_p}\psi(t,x)
\end{equation}
where $ L_P=\sqrt{G\hbar/c^3}$, $T_p=\sqrt{G\hbar/c^5}$,
$M_p=\sqrt{\hbar c/G}$ and $E_p=M_pc^2$. \\If we define
$N_m=m/M_p$ and $N_v=V/E_p$ we obtain for (\ref{ep5.2})
\begin{equation}
\label{ep5.3}
i T_p{\partial}_t\psi(t,x)=-\frac {L_p^2} {2N_m} {\partial}_x^2
\psi(t,x)+N_v\psi(t,x)
\end{equation}
By analogy with  ref.\cite{tt1} we define the fractional Schr\"{o}dinger equation
for all $\nu$ complex as:
\begin{equation}
\label{ep5.4}
(i T_p)^{\nu}{\partial}_t^{\nu}\psi(t,x)=-\frac {L_p^2} {2N_m} {\partial}_x^2
\psi(t,x)+N_v\psi(t,x)
\end{equation}
where the temporal fractionary derivative is defined following ref.\cite{tt2}
(see Appendix III)

\section{The Free Particle}

\setcounter{equation}{0}

From (\ref{ep5.4}) for the free particle the fractionary equation is:
\begin{equation}
\label{ep6.1}
(i{\partial}_t)^{\nu}\psi(t,x)+\frac {L_p^2} {2T_p^{\nu}N_m} {\partial}_x^2
\psi(t,x)=0
\end{equation}
By the use of the Fourier transform 
(complex in the temporal variable and real as usual in the spatial variable)
the corresponding equation is
(see Appendix II and ref.\cite{tt2})
\begin{equation}
\label{ep6.2}
\left(k_0^{\nu}-\frac {L_p^2} {2T_p^{\nu}N_m}k^2\right)
\hat{\psi}(k_0,k)=b(k_0,k)
\end{equation}
whose solution is:
\begin{equation}
\label{ep6.3}
\hat{\psi}(k_0,k)=\frac {b(k_0,k)} {k_0^{\nu}-\frac {L_p^2} {2T_p^{\nu}N_m}k^2}
\end{equation}
and in the configuration space (anti-transforming)
\begin{equation}
\label{ep6.4}
{\psi}(t,x)=
\oint\limits_{\Gamma}\int\limits_{-\infty}^{\infty}
\frac {a(k_0,k)} {k_0^{\nu}-\frac {L_p^2} {2T_p^{\nu}N_m}k^2}
e^{-i(k_0t+kx)}dk_0\;dk
\end{equation}
where:
\[a(k_0,k)=\frac {b(k_0,k)} {4{\pi}^2}\]
We proceed to analyze solutions of (\ref{ep6.4}) for some typical cases in the
following section.

\subsection*{Examples}

As a first example we consider the case $\nu=1/2$\\
Let $\alpha$ be given by: 
\begin{equation}
\label{ep6.5}
\alpha=\frac {L_p^2} {2 T_p^{\frac {1} {2}}N_m}
\end{equation}
From (\ref{ep6.4}) we obtain
\begin{equation}
\label{ep6.6}
{\psi}(t,x)=
\oint\limits_{\Gamma}\int\limits_{-\infty}^{\infty}
\frac {a(k_0,k)} {k_0^{\frac {1} {2}}- \alpha k^2}
e^{-i(k_0t+kx)}dk_0\;dk
\end{equation}
or equivalently:
\[\psi(t,x)=\int\limits_{-\infty}^{\infty}a(k)
e^{-i({\alpha}^2 k^4 t+kx)}dk + 
\int\limits_{-\infty}^0\int\limits_{-\infty}^{\infty}
a(k_0,k)\]
\begin{equation}
\label{ep6.7}
\left[\frac {1} {(k_0+i0)^{\frac {1} {2}}-\alpha k^2}-
\frac {1} {(k_0-i0)^{\frac {1} {2}}-\alpha k^2}\right]
e^{-i(k_0t+kx)}dk_0\;dk
\end{equation}
where:
\[a(k)=-4\pi i \alpha k^2 a(\alpha^2k^4,k)\]
With some of algebraic calculus we obtain for (\ref{ep6.7}):
\[\psi(t,x)=\int\limits_{-\infty}^{\infty}a(k)
e^{-i({\omega}^2 t+kx)}dk +\]
\begin{equation}
\label{ep6.8}
\int\limits_0^{\infty}\int\limits_{-\infty}^{\infty}
\frac {a(k_0,k)} {k_0+{\omega}^2}
e^{i(k_0t-kx)}dk_0\;dk
\end{equation}
with:
\[\omega=\alpha k^2\]
and where we have made the re-scaling:
\[-2ik_0^{\frac {1} {2}} a(-k_0,k)\rightarrow a(k_0,k)\]
The first term in (\ref{ep6.8}) represent free particle on-shell propagation
and the second term describes the contribution of off-shell modes.

As a second example we consider the case $\nu=1$.\\
In this case (\ref{ep6.4}) takes the form:
\begin{equation}
\label{ep6.9}
{\psi}(t,x)=
\oint\limits_{\Gamma}\int\limits_{-\infty}^{\infty}
\frac {a(k_0,k)} {k_0-\omega}
e^{-i(k_0t+kx)}dk_0\;dk
\end{equation}
Evaluating the integral in the variable $k_0$ we have:
\begin{equation}
\label{rp6.10}
\psi(t,x)=\int\limits_{-\infty}^{\infty}a(k)
e^{-i(\omega t+kx)} dk
\end{equation}
where $a(k)=-2\pi i a(\omega,k)$.
Thus we recover the usual expression for the free-particle wave function.

Finally we consider the case $\nu=2$. For it we have
\begin{equation}
\label{ep6.11}
{\psi}(t,x)=
\oint\limits_{\Gamma}\int\limits_{-\infty}^{\infty}
\frac {a(k_0,k)} {k_0^2-{\omega}^2}
e^{-i(k_0t+kx)}dk_0\;dk
\end{equation}
After to perform the integral in the variable $k_0$ we obtain from
(\ref{ep6.11}):
\begin{equation}
\label{ep6.12}
\psi(t,x)=\int\limits_{-\infty}^{\infty}
a(k) e^{-i(\omega t+kx)}+ b^+(k) e^{i(\omega t + kx)}dk
\end{equation}
with $a(k)=-2\pi i a(\omega,k)$ and $b^+(k)=-2\pi i a(-\omega,-k)$

\section{The Potential Well}

\setcounter{equation}{0}

We consider in this section the potential well. The fractionary
equation for a particle confined to move within interval $0\leq x\leq a$ is:
\begin{equation}
\label{ep7.1}
(iT_p)^{\nu} {\partial}_t^{\nu}\psi(t,x)=-
\frac {L_p^2} {2N_m}{\partial}_x^2\psi(t,x)
\end{equation}
To solve this equation we use the method of separation of variables.
Thus if we write:
\begin{equation}
\label{ep7.2}
\psi(t,x)={\psi}_1(t){\psi}_2(x)
\end{equation}
As is usual we obtain:
\begin{equation}
\label{ep7.3}
\frac {(iT_p)^{\nu} {\partial}_t^{\nu}\psi_1(t)} {\psi_1(t)}=-
\frac {\frac {L_p^2} {2N_m}{\partial}_x^2\psi_2(x)} {\psi_2(x)}=
\lambda
\end{equation}
Then we conclude that $\psi_2(x)$ satisfies:
\begin{equation}
\label{ep7.4}
{\partial}_x^2{\psi}_2(x)+\frac {2\lambda N_m} {L_p^2}
{\psi}_2(x)=0
\end{equation}
The solution of (\ref{ep7.4}) is the habitual one:
\begin{equation}
\label{ep7.5}
{\psi}_{2n}(x)=b_n\sin\left(\frac {n\pi} {a}x\right)
\end{equation}
with:
\begin{equation}
\label{ep7.6}
\lambda_n=\frac {1} {2N_m}
\left(\frac {n\pi L_p} {a}\right)^2
\end{equation}
and the boundary  conditions satisfied by $\psi_{2n}(x)$ are:
\[\psi_{2n}(0)=\psi_{2n}(a)=0\]
As a consequence of (\ref{ep7.3}),(\ref{ep7.5}) and (\ref{ep7.6})
the Fourier transform ${\hat{\psi}_1}(k_0)$ of $\psi_1(t)$
should be satisfy:
\begin{equation}
\label{ep7.7}
(k_0^{\nu}-\lambda_n){\hat{\psi}}_{1n}(k_0)=0
\end{equation}
whose solution is:
\begin{equation}
\label{ep7.8}
{\hat{\psi}}_{1n}(k_0)=\frac {c_n(k_0)} {k_0^\nu-\lambda_n}
\end{equation}
Therefore the final general solution  for $\psi(t,x)$ is:
\begin{equation}
\label{ep7.9}
\psi(t,x)=\sum\limits_{n=1}^{\infty}\sin\left(\frac {n\pi} {a}x\right)
\oint\limits_{\Gamma} \frac {a_n(k_0)e^{-ik_0t}} {k_0^\nu-\lambda_n}
dk_0
\end{equation}
where we have defined:
\[a_n(k_0)=\frac {b_nc_n(k_0)} {2\pi}\]
which is an entire analytic function of $k_0$.

\subsection*{Examples}

As a first example we consider the case $\nu=1/2$. 
For it the solution (\ref{ep7.9}) takes the form:
\begin{equation}
\label{ep7.10}
\psi(t,x)=\sum\limits_{n=1}^{\infty}\sin\left(\frac {n\pi} {a}x\right)
\oint\limits_{\Gamma} \frac {a_n(k_0)e^{-ik_0t}} 
{k_0^{\frac {1} {2}}-\lambda_n}
dk_0
\end{equation}
or equivalently:
\[\psi(t,x)=\sum\limits_{n=1}^{\infty}a_n
\sin\left(\frac {n\pi} {a} x\right)
e^{-i\lambda_n^2 t}+\]
\[\sum\limits_{n=1}^{\infty}\sin\left(\frac {n\pi} {a} x\right)
\int\limits_{-\infty}^0\left[\frac {1} {(k_0+i0)^{\frac {1} {2}}-\lambda_n}-
\frac {1} {(k_0-i0)^{\frac {1} {2}}-\lambda_n}\right]\;\times\]
\begin{equation}
\label{ep7.11}
a_n(k_0)e^{-ik_0 t}\;dk_0
\end{equation}
After performing some algebra we have for (\ref{ep7.11}) the expression:
\[\psi(t,x)=\sum\limits_{n=1}^{\infty}a_n
\sin\left(\frac {n\pi} {a} x\right)
e^{-i\lambda_n^2 t}+\]
\begin{equation}
\label{ep7.12}
\sum\limits_{n=1}^{\infty}\sin\left(\frac {n\pi} {a} x\right)
\int\limits_0^{\infty}\frac {a_n(k_0)} {k_0+\lambda_n^2}
e^{-ik_0 t}\;dk_0
\end{equation}
Analogously as before, the second term in (\ref{ep7.12}) represents 
of-shell stationary modes.

As a second example we consider $\nu=1$.
In this case:
\begin{equation}
\label{ep7.13}
\psi(t,x)=\sum\limits_{n=1}^{\infty}\sin\left(\frac {n\pi} {a}x\right)
\oint\limits_{\Gamma} \frac {a_n(k_0)e^{-ik_0t}} {k_0-\lambda_n}
dk_0
\end{equation}
Performing the integral in the variable $k_0$ we have:
\begin{equation}
\label{ep7.14}
\psi(t,x)=\sum\limits_{n=1}^{\infty}a_n
\sin\left(\frac {n\pi} {a} x\right)
e^{-i\lambda_n t}
\end{equation}
Which is the familiar general solution for the infinite well.

Finally for $\nu=2$:
\begin{equation}
\label{ep7.15}
\psi(t,x)=\sum\limits_{n=1}^{\infty}\sin\left(\frac {n\pi} {a}x\right)
\oint\limits_{\Gamma} \frac {a_n(k_0)e^{-ik_0t}} {k_0^2-\lambda_n}
dk_0
\end{equation}
and after to compute the integral:
\begin{equation}
\label{ep7.16}
\psi(t,x)=\sum\limits_{n=1}^{\infty}
\sin\left(\frac {n\pi} {a} x\right)\left(a_n
e^{-i\sqrt{\lambda_n} t}+b_ne^{+i\sqrt{\lambda_n}t}\right) 
\end{equation}
with $a_n=a_n(\sqrt{{\lambda}_n})$ and $b_n^+=a_n(-\sqrt{{\lambda}_n})$

\section{The Green Function for The Free Particle}

\setcounter{equation}{0}

As other application that shows the generality of the fractional
calculus \\defined with the use of ultradistributions, we give the
evaluation of the Green function corresponding to the
free particle. Let $\beta$ be defined as:
\begin{equation}
\label{ep8.1}
\beta^2=\frac {L_p^2} {2 T_p^{\nu}N_m}
\end{equation}
Then $G(t-t^{'},x-x^{'})$ should be satisfy the equation:
\begin{equation}
\label{ep8.2}
(i\partial_t)^{\nu}G(t-t^{'},x-x^{'})+\beta^2{\partial}_x^2
G(t-t^{'},x-x^{'})=\delta(t-t^{'})\delta(x-x^{'})
\end{equation}
As $G$ is function of $(t-t^{'},x-x^{'})$ it is sufficient to
consider $G$ as function of $(t,x)$:
\begin{equation}
\label{ep8.3}
(i\partial_t)^{\nu}G(t,x)+\beta^2{\partial}_x^2
G(t,x)=\delta(t)\delta(x)
\end{equation}
For the Fourier transform $\hat{G}$ of $G$ we have:
\begin{equation}
\label{ep8.4}
(k_0^{\nu}-\beta^2k^2)\hat{G}(k_0,k)=\frac {Sgn[\Im(k_0)]} {2}
+a(k_0,k)
\end{equation}
where $a(k_0,k)$ is as usual a 
rapidly decreasing analytic entire function of the
variable $k_0$ . Selecting:
\[a(k_0,k)=\frac {1} {2}\]
we obtain the equation for the retarded Green function:
\begin{equation}
\label{ep8.5}
(k_0^{\nu}-\beta^2k^2){\hat{G}}_{ret}(k_0,k)=H[\Im(k_0)]
\end{equation}
and then:
\begin{equation}
\label{ep8.6}
G_{ret}(t,x)=\frac {1} {4\pi^2}\oint\limits_{\Gamma}
\int\limits_{-\infty}^{\infty}
\frac {H[\Im(k_0)]} {k_0^{\nu}-\beta^2k^2}
e^{-i(k_0t+kx)}\;dk_0\;dk
\end{equation}
If we take:
\[a(k_0,k)=-\frac {1} {2}\]
we obtain the advanced Green function:
\begin{equation}
\label{ep8.7}
G_{adv}(t,x)=-\frac {1} {4\pi^2}\oint\limits_{\Gamma}
\int\limits_{-\infty}^{\infty}
\frac {H[-\Im(k_0)]} {k_0^{\nu}-\beta^2k^2}
e^{-i(k_0t+kx)}\;dk_0\;dk
\end{equation}
For the Wheeler Green function (half advanced plus half retarded):
\begin{equation}
\label{ep8.8}
G_W(t,x)=\frac {1} {2} [G_{adv}(t,x)+G_{ret}(t,x)]
\end{equation}
we have:
\begin{equation}
\label{ep8.9}
G_W(t,x)=\frac {1} {8\pi^2}\oint\limits_{\Gamma}
\int\limits_{-\infty}^{\infty}
\frac {Sgn[\Im(k_0)]} {k_0^{\nu}-\beta^2k^2}
e^{-i(k_0t+kx)}\;dk_0\;dk
\end{equation}

\subsection*{Example}

When we select $\nu=1$ we obtain the usual Green functions
of Quantum Mechanics. For example for $G_{ret}$ we have:
\begin{equation}
\label{ep8.10}
G_{ret}(t,x)=\frac {1} {4\pi^2}\oint\limits_{\Gamma}
\int\limits_{-\infty}^{\infty}
\frac {H[\Im(k_0)]} {k_0-\beta^2k^2}
e^{-i(k_0t+kx)}\;dk_0\;dk
\end{equation}
or equivalently:
\begin{equation}
\label{ep8.11}
G_{ret}(t,x)=\frac {1} {4\pi^2}\int\limits_{-\infty}^{\infty}
\int\limits_{-\infty}^{\infty}
\frac {1} {(k_0+i0)-\beta^2k^2}
e^{-i(k_0t+kx)}\;dk_0\;dk
\end{equation}
After the evaluation of the integral in the variable $k_0$,
$G_{ret}$ takes the form:
\begin{equation}
\label{ep8.12}
G_{ret}(t,x)=-\frac {i} {2\pi}H(t)\int\limits_{-\infty}^{\infty}
e^{-i(\beta^2k^2t+kx)}\;dk
\end{equation}
With a square's completion (\ref{ep8.12}) transforms into:
\begin{equation}
\label{ep8.13}
G_{ret}(t,x)=-\frac {iH(t)} {2\pi\beta\sqrt{t}}
e^{\frac {ix^2} {4\beta^2t}}\int\limits_{-\infty}^{\infty}
e^{is^2}ds
\end{equation}
From the result of ref.\cite{tt3}
\begin{equation}
\label{ep8.14}
\int\limits_{-\infty}^{\infty}
e^{is^2}ds=\sqrt{\pi}e^{-i\frac {\pi} {4}}
\end{equation}
we have
\begin{equation}
\label{ep8.15}
G_{ret}(t,x)=-iH(t)\left(\frac {m} {2\pi i \hbar t}\right)^{\frac {1} {2}}
e^{\frac {imx^2} {2\hbar t}}
\end{equation}
Taking into account that for $\nu=1$:
\[\beta^2=\frac {\hbar} {2m}\]
we obtain the usual form of $G_{ret}$ (see ref.\cite{tt4})
\begin{equation}
\label{ep8.16}
G_{ret}(t-t^{'},x-x^{'})=-iH(t-t^{'})\left(\frac {m} {2\pi i \hbar (t-t^{'})}\right)^{\frac {1} {2}}
e^{\frac {im(x-x^{'})^2} {2\hbar (t-t^{'})}}
\end{equation}
With a similar calculus we have for $G_{adv}$:
\begin{equation}
\label{ep8.17}
G_{adv}(t-t^{'},x-x^{'})=iH(t^{'}-t)\left(\frac {m} {2\pi i \hbar (t^{'}-t)}\right)^{\frac {1} {2}}
e^{\frac {im(x-x^{'})^2} {2\hbar (t-t^{'})}}
\end{equation}
and for $G_W$:
\begin{equation}
\label{ep8.18}
G_W(t-t^{'},x-x^{'})=
-\frac {i} {2} Sgn(t-t^{'})\left(\frac {m} {2\pi i \hbar |t-t^{'}|}\right)^{\frac {1} {2}}
e^{\frac {im(x-x^{'})^2} {2\hbar (t-t^{'})}}
\end{equation}

\section{Discussion}

In a earlier paper (ref.\cite{tt2} we have shown the existence of 
a general fractional calculus defined via tempered ultradistributions. 
All ultradistributions provide integrands that are
analytic functions along the integration path. 
These  properties 
show that tempered ultradistributions provide an
appropriate framework for applications to fractional calculus.
With the use of this calculus we have generalized in the present work
the results obtained by Naber (ref.\cite{tt1}).
We have defined the fractionary Schr\"{o}dinger 
equation for all values of the
complex variable $\nu$ and treated the cases of the 
free particle and the potential well. For $\nu=1$ the results 
obtained coincide with the usual Quantum Mechanics,
and the cases $\nu=1/2$ and $\nu=2$ have shown the appearance 
of extra terms, besides to those with the usual ($\nu=1$) framework.
We have obtained a general expression for the
Green function of the free particle and shown that for $\nu=1$
this Green function coincide with the obtained in ref.\cite{tt4}..
For the benefit of the reader
we give in  this paper two Appendixes with the main characteristics
of n-dimensional tempered ultradistributions and their Fourier
anti-transformed distributions of the exponential type, and a third 
Appendix about the general fractional calculus defined via the use
of tempered ultradistributions.

\section{Appendix I: Distributions of Exponential Type}

\setcounter{equation}{0}

For the sake of the reader we shall present a brief description of the
principal properties of Tempered Ultradistributions.

{\bf Notations}.
The notations are almost textually taken from ref\cite{tp2}.
Let $\boldsymbol{{\mathbb{R}}^n}$
(res. $\boldsymbol{{\mathbb{C}}^n}$) be the real (resp. complex)
n-dimensional space whose points are denoted by $x=(x_1,x_2,...,x_n)$
(resp $z=(z_1,z_2,...,z_n)$). We shall use the notations:

(i) $x+y=(x_1+y_1,x_2+y_2,...,x_n+y_n)$\; ; \;
    $\alpha x=(\alpha x_1,\alpha x_2,...,\alpha x_n)$

(ii)$x\geqq 0$ means $x_1\geqq 0, x_2\geqq 0,...,x_n\geqq 0$

(iii)$x\cdot y=\sum\limits_{j=1}^n x_j y_j$

(iV)$\mid x\mid =\sum\limits_{j=1}^n \mid x_j\mid$

Let $\boldsymbol{{\mathbb{N}}^n}$ be the set of n-tuples of natural
numbers. If $p\in\boldsymbol{{\mathbb{N}}^n}$, then
$p=(p_1, p_2,...,p_n)$,
and $p_j$ is a natural number, $1\leqq j\leqq n$. $p+q$ denote
$(p_1+q_1, p_2+q_2,..., p_n+q_n)$ and $p\geqq q$ means $p_1\geqq q_1,
p_2\geqq q_2,...,p_n\geqq q_n$. $x^p$ means $x_1^{p_1}x_2^{p_2}...
x_n^{p_n}$. We shall denote by
$\mid p\mid=\sum\limits_{j=1}^n  p_j $ and by $D^p$ we denote the
differential operator ${\partial}^{p_1+p_2+...+p_n}/\partial{x_1}^{p_1}
\partial{x_2}^{p_2}...\partial{x_n}^{p_n}$

For any natural $k$ we define $x^k=x_1^k x_2^k...x_n^k$
and ${\partial}^k/\partial x^k=
{\partial}^{nk}/\partial x_1^k\partial x_2^k...\partial x_n^k$

The space $\boldsymbol{{\cal H}}$  of test functions
such that $e^{p|x|}|D^q\phi(x)|$ is bounded for any p and q
is defined ( ref.\cite{tp2} ) by means
of the countably set of norms:
\begin{equation}
\label{ep2.1}
{\|\hat{\phi}\|}_p=\sup_{0\leq q\leq p,\,x}
e^{p|x|} \left|D^q \hat{\phi} (x)\right|\;\;\;,\;\;\;p=0,1,2,...
\end{equation}
According to reference\cite{tp5} $\boldsymbol{{\cal H}}$  is a
$\boldsymbol{{\cal K}\{M_p\}}$ space
with:
\begin{equation}
\label{ep2.2}
M_p(x)=e^{(p-1)|x|}\;\;\;,\;\;\; p=1,2,...
\end{equation}
$\boldsymbol{{\cal K}\{e^{(p-1)|x|}\}}$ satisfies condition
$\boldsymbol({\cal N})$
of Guelfand ( ref.\cite{tp4} ). It is a countable Hilbert and nuclear
space:
\begin{equation}
\label{ep2.3}
\boldsymbol{{\cal K}\{e^{(p-1)|x|}\}} =\boldsymbol{{\cal H}} =
\bigcap\limits_{p=1}^{\infty}\boldsymbol{{\cal H}_p}
\end{equation}
where $\boldsymbol{{\cal H}_p}$ is obtained by completing
$\boldsymbol{{\cal H}}$ with the norm induced by
the scalar product:
\begin{equation}
\label{ep2.4}
{<\hat{\phi}, \hat{\psi}>}_p = \int\limits_{-\infty}^{\infty}
e^{2(p-1)|x|} \sum\limits_{q=0}^p D^q \overline{\hat{\phi}} (x) D^q
\hat{\psi} (x)\;dx \;\;\;;\;\;\;p=1,2,...
\end{equation}
where $dx=dx_1\;dx_2...dx_n$

If we take the usual scalar product:
\begin{equation}
\label{ep2.5}
<\hat{\phi}, \hat{\psi}> = \int\limits_{-\infty}^{\infty}
\overline{\hat{\phi}}(x) \hat{\psi}(x)\;dx
\end{equation}
then $\boldsymbol{{\cal H}}$, completed with (\ref{ep2.5}), is the Hilbert space
$\boldsymbol{H}$
of square integrable functions.

The space of continuous linear functionals defined on
$\boldsymbol{{\cal H}}$ is the space
$\boldsymbol{{\Lambda}_{\infty}}$ of the distributions of the exponential
type ( ref.\cite{tp2} ).

The ``nested space''
\begin{equation}
\label{ep2.6}
{\Large{H}}=
\boldsymbol{(}\boldsymbol{{\cal H}},\boldsymbol{H},
\boldsymbol{{\Lambda}_{\infty}} \boldsymbol{)}
\end{equation}
is a Guelfand's triplet ( or a Rigged Hilbert space \cite{tp4} ).

In addition we have: $\boldsymbol{{\cal H}}\subset\boldsymbol{{\cal S}}
\subset\boldsymbol{H}\subset\boldsymbol{{\cal S}^{'}}\subset
\boldsymbol{{\Lambda}_{\infty}}$, where $\boldsymbol{{\cal S}}$ is the
Schwartz space of rapidly decreasing test functions (ref\cite{tp6}).

Any Guelfand's triplet
${\Large{G}}=\boldsymbol{(}\boldsymbol{\Phi},
\boldsymbol{H},\boldsymbol{{\Phi}^{'}}\boldsymbol{)}$
has the fundamental property that a linear and symmetric operator
on $\boldsymbol{\Phi}$, admitting an extension to a self-adjoint
operator in
$\boldsymbol{H}$, has a complete set of generalized eigen-functions
in $\boldsymbol{{\Phi}^{'}}$ with real eigenvalues.

\section{Appendix II: Tempered Ultradistributions}
\setcounter{equation}{0}

The Fourier transform of a function $\hat{\phi}\in \boldsymbol{{\cal H}}$
is
\begin{equation}
\label{ep3.1}
\phi(z)=\frac {1} {2\pi}
\int\limits_{-\infty}^{\infty}\overline{\hat{\phi}}(x)\;e^{iz\cdot x}\;dx
\end{equation}
$\phi(z)$ is entire analytic and rapidly decreasing on straight lines
parallel
to the real axis. We shall call $\boldsymbol{{\EuFrak H}}$
the set of all such functions.
\begin{equation}
\label{ep3.2}
\boldsymbol{{\EuFrak H}}={\cal F}\left\{\boldsymbol{{\cal H}}\right\}
\end{equation}
It is a $\boldsymbol{{\cal Z}\{M_p\}}$ space ( ref.\cite{tp5} ),
countably normed and complete, with:
\begin{equation}
\label{ep3.3}
M_p(z)= (1+|z|)^p
\end{equation}
$\boldsymbol{{\EuFrak H}}$ is also a nuclear space with norms:
\begin{equation}
\label{ep3.4}
{\|\phi\|}_{pn} = \sup_{z\in V_n} {\left(1+|z|\right)}^p
|\phi (z)|
\end{equation}
where $V_k=\{z=(z_1,z_2,...,z_n)\in\boldsymbol{{\mathbb{C}}^n}:
\mid Im z_j\mid\leqq k, 1\leqq j \leqq n\}$

We can define the usual scalar product:
\begin{equation}
\label{ep3.5}
<\phi (z), \psi (z)>=\int\limits_{-\infty}^{\infty}
\phi(z) {\psi}_1(z)\;dz =
\int\limits_{-\infty}^{\infty} \overline{\hat{\phi}}(x)
\hat{\psi}(x)\;dx
\end{equation}
where:
\[{\psi}_1(z)=\int\limits_{-\infty}^{\infty}
\hat{\psi}(x)\; e^{-iz\cdot x}\;dx\]
and $dz=dz_1\;dz_2...dz_n$

By completing $\boldsymbol{{\EuFrak H}}$ with the norm induced by (\ref{ep3.5})
we get the Hilbert space of square integrable functions.

The dual of $\boldsymbol{{\EuFrak H}}$ is the space
$\boldsymbol{{\cal U}}$ of tempered ultradistributions
( ref.\cite{tp2} ). In other words, a tempered ultradistribution is
a continuous linear functional defined on the space
$\boldsymbol{{\EuFrak H}}$ of entire
functions rapidly decreasing on straight lines parallel to the real axis.

The set
${\Large{U}}=
\boldsymbol{({\EuFrak H},H,{\cal U})}$ is also a Guelfand's triplet.

Moreover, we have: $\boldsymbol{{\EuFrak H}}\subset\boldsymbol{{\cal S}}
\subset\boldsymbol{H}\subset\boldsymbol{{\cal S}^{'}}\subset
\boldsymbol{{\cal U}}$.

$\boldsymbol{{\cal U}}$ can also be characterized in the following way
( ref.\cite{tp2} ): let $\boldsymbol{{\cal A}_{\omega}}$ be the space of
all functions $F(z)$ such that:

${\Large {\boldsymbol{I}}}$-
$F(z)$ is analytic for $\{z\in \boldsymbol{{\mathbb{C}}^n} :
|Im(z_1)|>p, |Im(z_2)|>p,...,|Im(z_n)|>p\}$.

${\Large {\boldsymbol{II}}}$-
$F(z)/z^p$ is bounded continuous  in
$\{z\in \boldsymbol{{\mathbb{C}}^n} :|Im(z_1)|\geqq p,|Im(z_2)|\geqq p,
...,|Im(z_n)|\geqq p\}$,
where $p=0,1,2,...$ depends on $F(z)$.

Let $\boldsymbol{\Pi}$ be the set of all $z$-dependent pseudo-polynomials,
$z\in \boldsymbol{{\mathbb{C}}^n}$.
Then $\boldsymbol{{\cal U}}$ is the quotient space:

${\Large {\boldsymbol{III}}}$-
$\boldsymbol{{\cal U}}=\boldsymbol{{\cal A}_{\omega}/\Pi}$

By a pseudo-polynomial we understand a function of $z$ of the form $\;\;$
$\sum_s z_j^s G(z_1,...,z_{j-1},z_{j+1},...,z_n)$ with
$G(z_1,...,z_{j-1},z_{j+1},...,z_n)\in\boldsymbol{{\cal A}_{\omega}}$

Due to these properties it is possible to represent any ultradistribution
as ( ref.\cite{tp2} ):
\begin{equation}
\label{ep3.6}
F(\phi)=<F(z), \phi(z)>=\oint\limits_{\Gamma} F(z) \phi(z)\;dz
\end{equation}
$\Gamma={\Gamma}_1\cup{\Gamma}_2\cup ...{\Gamma}_n$
where the path ${\Gamma}_j$ runs parallel to the real axis from
$-\infty$ to $\infty$ for $Im(z_j)>\zeta$, $\zeta>p$ and back from
$\infty$ to $-\infty$ for $Im(z_j)<-\zeta$, $-\zeta<-p$.
( $\Gamma$ surrounds all the singularities of $F(z)$ ).

Formula (\ref{ep3.6}) will be our fundamental representation for a tempered
ultradistribution. Sometimes use will be made of ``Dirac formula''
for ultradistributions ( ref.\cite{tp1} ):
\begin{equation}
\label{ep3.7}
F(z)=\frac {1} {(2\pi i)^n}\int\limits_{-\infty}^{\infty}
\frac {f(t)} {(t_1-z_1)(t_2-z_2)...(t_n-z_n)}\;dt
\end{equation}
where the ``density'' $f(t)$ is such that
\begin{equation}
\label{ep3.8}
\oint\limits_{\Gamma} F(z) \phi(z)\;dz =
\int\limits_{-\infty}^{\infty} f(t) \phi(t)\;dt
\end{equation}
While $F(z)$ is analytic on $\Gamma$, the density $f(t)$ is in
general singular, so that the r.h.s. of (\ref{ep3.8}) should be interpreted
in the sense of distribution theory.

Another important property of the analytic representation is the fact
that on $\Gamma$, $F(z)$ is bounded by a power of $z$ ( ref.\cite{tp2} ):
\begin{equation}
\label{ep3.9}
|F(z)|\leq C|z|^p
\end{equation}
where $C$ and $p$ depend on $F$.

The representation (\ref{ep3.6}) implies that the addition of a
pseudo-polynomial $P(z)$ to $F(z)$ do not alter the ultradistribution:
\[\oint\limits_{\Gamma}\{F(z)+P(z)\}\phi(z)\;dz=
\oint\limits_{\Gamma} F(z)\phi(z)\;dz+\oint\limits_{\Gamma}
P(z)\phi(z)\;dz\]
But:
\[\oint\limits_{\Gamma} P(z)\phi(z)\;dz=0\]
as $P(z)\phi(z)$ is entire analytic in some of the variables $z_j$
( and rapidly decreasing ),
\begin{equation}
\label{ep3.10}
\therefore \;\;\;\;\oint\limits_{\Gamma} \{F(z)+P(z)\}\phi(z)\;dz=
\oint\limits_{\Gamma} F(z)\phi(z)\;dz
\end{equation}

\section{Appendix III: Fractional Calculus}

\setcounter{equation}{0}
The purpose of this sections is to introduce definition of fractional
derivation and integration given in ref. \cite{tp1}.
This definition unifies the notion of integral and derivative in one only
operation.
Let $\hat{f}(x)$ a distribution of exponential type and $F(\Omega)$
the complex Fourier transformed Tempered Ultradistribution.
Then:
\begin{equation}
\label{ep4.1}
F(k)=H[\Im(k)]\int\limits_0^{\infty}\hat{f}(x) e^{ik x}\;dx-
H[-\Im(k)]\int\limits_{-\infty}^0\hat{f}(x) e^{ik x}\;dx
\end{equation}
($H(x)$ is the Heaviside step function)
and
\begin{equation}
\label{ep4.2}
\hat{f}(x)=\frac {1} {2\pi}\oint\limits_{\Gamma}F(k)
e^{-ik x}\;dk
\end{equation}
where the contour $\Gamma$ surround all singularities of
$F(k)$ and runs parallel to real axis from $-\infty$ to
$\infty$ above the real axis and from $\infty$ to $-\infty$
below the real axis.
According to \cite{tp1} the fractional derivative of $\hat{f}(x)$ is given by
\begin{equation}
\label{ep4.3}
\frac {d^{\lambda}\hat{f}(x)} {dx^{\lambda}}=\frac {1} {2\pi}\oint\limits_{\Gamma}
(-ik)^{\lambda}
F(k)
e^{-ik x}\;dk+
\oint\limits_{\Gamma}(-ik)^{\lambda}a(k)
e^{-ik x}\;dk
\end{equation}
Where $a(k)$ is entire analytic and rapidly decreasing.
If $\lambda=-1$, $d^{\lambda}/dx^{\lambda}$ is the inverse of the derivative
(an integration). In this case the second term of the right side of (\ref{ep4.3})
gives a primitive of $\hat{f}(x)$. Using Cauchy's theorem the additional
term is 
\begin{equation}
\label{ep4.4}
\oint \frac {a(k)} {k}e^{-ik x} dk=
2\pi a(0)
\end{equation}
Of course, an integration should give a primitive plus an arbitrary constant.
Analogously when $\lambda=-2$ (a double iterated integration) we have
\begin{equation}
\label{ep4.5}
\oint \frac {a(k)} {{k}^2}e^{-ik x} dk=
\gamma+\delta x
\end{equation}
where $\gamma$ and $\delta$ are arbitrary constants.

\end{document}